\documentclass[reprint,aps,preprintnumbers,amsmath,amssymb,nofootinbib,epsf,epsfig]
{revtex4-2}
\usepackage[utf8]{inputenc}
\usepackage{graphicx}
\usepackage{amsmath}
\usepackage{amssymb}
\usepackage{array}
\usepackage{pifont}
\usepackage{float}
\usepackage{booktabs}
\usepackage{multirow}
\usepackage{color}
\usepackage{caption}
\usepackage{subcaption}
\usepackage{natbib}
\usepackage{pifont}
\usepackage[colorlinks, citecolor=cyan]{hyperref}
\usepackage{epsfig}
\usepackage{slashed}
\usepackage{appendix}
\usepackage{comment}

\def\s1{\hat s}

\newcommand{\be}{\begin{equation}}
\newcommand{\ee}{\end{equation}}
\newcommand{\bea}{\begin{eqnarray}}
\newcommand{\eea}{\end{eqnarray}}
\newcommand{\nn}{\nonumber}

\def\s1{\hat s}

\begin{document}
\title{\large Signature of (axial)vector operators in $B_c\to D_s^{(*)} \mu^+ \mu^-$ decays}

\author{Manas Kumar Mohapatra$^a$} 
\email{manasmohapatra12@gmail.com} 
\author{Ajay Kumar Yadav$^b$}
\email{yadavajaykumar286@gmail.com}
\author{Suchismita Sahoo$^b$}
\email{suchismita8792@gmail.com}
\affiliation{
School of Physics, University of Hyderabad, Hyderabad-500046, India $^a$ \\ Department of Physics, Central University of Karnataka, Kalaburagi-585367, India $^b$}
\begin{abstract}
The persistent anomalies observed in $B$ mesons related to $b \to s \mu^+ \mu^-$ transitions point to the potential for new physics beyond the Standard Model. To explore this further, we conduct an analysis of  the exclusive semileptonic decays  $B_c\to D_s^{(*)} \mu^+ \mu^-$ within  an effective field theory formalism. We perform a global fit to new (axial)vector couplings using the current experimental data for $b \to s \mu^+ \mu^-$ transitions. We compute the branching ratios and physical observables such as forward-backward asymmetries, lepton polarization asymmetries, form factor independent observables for  $B_c\to D_s^{(*)} \mu^+ \mu^-$,  both within the Standard Model and in new physics scenarios. Additionally, we examine the existence of lepton flavor universality violation in the $B_c\to D_s^{(*)} \ell^+ \ell^-$ process. 
\end{abstract}
\maketitle
\section{Introduction}
Rare $B$-meson decays mediated by the $b \to s \mu^+\mu^-$ transition have emerged as one of the most promising probes of new physics (NP) beyond the Standard Model (SM). The motivation to study these channels can be summarized as follows:  

\begin{itemize}
    \item \textbf{FCNC nature:} The $b \to s \mu^+ \mu^-$ transition is a flavour-changing neutral current (FCNC)~\cite{LHCb:2015svh,LHCb:2020dof,LHCb:2020gog,LHCb:2020lmf,PhysRevLett.131.051803,Belle,LHCb:2023lyb,Capdevila:2023yhq,Sahoo:2016edx,Hurth:2023jwr,Mahmoudi:2024zna,Rajeev:2020aut,Rajeev:2021ntt,Mohapatra:2021izl,Das:2023kch,MunirBhutta:2020ber,SinghChundawat:2022ldm,Yadav:2024cbt,Albrecht:2021tul,London:2021lfn}, forbidden at tree level and occurring only via loops, making it highly sensitive to NP.  
    \item \textbf{Anomalies:} Deviations from SM predictions have been observed in angular observables, e.g.\ the $P'_5$ observable in $B^0 \to K^{*}\mu^+\mu^-$ decays shows $3.3\sigma$ and $2.1\sigma$ deviations in the bins $q^2 \in [4,6]$ and $[4,8]$, respectively~\cite{LHCb:2020lmf,Belle:2016xuo}.  
    \item \textbf{Lepton non-universality:} Ratios such as $R_{K^{(*)}}$~\cite{LHCb:2022vje} in $q^2\in [0.1,1.1]$ and $[1.1,6]$, along with the isospin partners $R_{K_S}$ ($R_{K^{*+}}$)~\cite{LHCb:2021lvy}, exhibit tensions of $0.2\sigma$ and $1.4\sigma$ ($1.5\sigma$) with SM expectations.  
    \item \textbf{Branching ratios:} Discrepancies in the differential branching fractions of $B \to K^{+}\mu^+\mu^-$~\cite{LHCb:2022vje} and $B_s \to \phi \mu^+\mu^-$~\cite{LHCb:2021zwz} have been reported at the level of $4.2\sigma$ and $3.6\sigma$, respectively.  
    \item \textbf{Global fits:} Combined analyses of $b \to s \mu^+\mu^-$ data suggest coherent NP patterns, particularly in the vector and axial-vector Wilson coefficients (WCs)~\cite{Altmannshofer:2021qrr,Descotes-Genon:2015uva}.  
\end{itemize}

Beyond these benchmark channels, several other mesonic ($B \to K^{(*)}\ell^+\ell^-$, $B_s \to \phi \ell^+\ell^-$) and baryonic ($\Lambda_b \to \Lambda \ell^+\ell^-$) decays undergoing $b \to s \ell^+ \ell^-$ transitions~\cite{Blake:2019guk,LHCb:2018jna,Li:2011nf,Rajeev:2020aut,Mohapatra:2021izl} provide complementary probes, where similar NP effects may also manifest. In this context, the $B_c$ meson, consisting of both $b$ and $c$ quarks, offers a unique opportunity: compared to other $B$ mesons, it provides a broader kinematic phase space and richer decay dynamics. The qualitative features of the predictions of the decay observables resemble those observed in $B \to (K^{(*)}, \phi)\mu^+\mu^-$. Additionally, the change in the spectator quark modifies both the hadronic form factors, leading to quantitative differences in branching fractions and angular observables.  From the theoretical side, the decays $B_c \to D_s^{(*)}\mu^+\mu^-$, mediated by FCNC $b \to s \mu^+\mu^-$ transitions, have been investigated in various approaches, including the covariant confined quark model~\cite{Ivanov:2019nqd,Ivanov:2024iat}, light-front quark model~\cite{Choi:2010ha}, QCD sum rules~\cite{Azizi:2008vv,Azizi:2008vy,Cooper:2021bkt}, and relativistic quark model~\cite{Ebert:2010dv,Zaki:2023mcw,Mohapatra:2021ynn}. Beyond the SM, both model-independent analyses~\cite{Dutta:2019wxo,Li:2023mrj,Ju:2013oba} and specific NP scenarios such as supersymmetry~\cite{Ahmed:2011sa}, leptoquarks, and $Z'$ bosons~\cite{Mohapatra:2021ynn,Lu:2012qnh} have been studied.

On the experimental side, $B_c$ decays became accessible after the first observation of the $B_c$ meson at CDF through $B_c \to J/\psi \ell \nu$~\cite{CDF:1998ihx}. The $B_c$ fragmentation fraction, $f_c$, was measured by LHCb to be nearly three orders of magnitude smaller than that of $B_u$ mesons, $f_u$~\cite{LHCb:2019tea}, making their detection challenging. For example, an upper bound has been placed on $B_c^+ \to D_s^+\ell^+\ell^-$ as $(f_c/f_u)\times\mathcal{B}(B_c^+\to D_s^+\ell^+\ell^-)<9.6\times10^{-8}$~\cite{LHCb:2023lyb}. Although the predicted branching fractions of $B_c\to D_s^{*}\ell^+\ell^-$ are comparable to those of $B_s$ decays, reconstruction is complicated by the additional decay $D_s^* \to D_s\gamma$, especially with current luminosities~\cite{LHCb:2025uci}. Nevertheless, future LHC runs are expected to produce $10^8$--$10^{10}$ $B_c$ mesons~\cite{Chang:1992jb,Cheung:1993qi,Braaten:1993jn,Stone:1997vk}, corresponding to up to $2\times 10^{10}$ events per year at $\mathcal{L}=10^{34}\,\text{cm}^{-2}\text{s}^{-1}$~\cite{Gouz:2002kk,PepeAltarelli:2008yyl}, providing excellent prospects for investigating these rare decays. The forthcoming measurements of $B_c \to D_s^{(*)}\mu^+\mu^-$ at LHC will allow these decays to serve as independent probes of the anomalies observed in $b \to s \mu^+ \mu^-$ processes, thus providing a valuable cross-check of NP scenarios.

Motivated by these theoretical and experimental considerations, in this work we study $B_c \to D_s^{(*)}\mu^+\mu^-$ decays in a model-independent effective field theory (EFT) framework. We focus on NP contributions to vector and axial-vector operators, which directly impact anomalies in branching ratios, angular observables, and lepton flavor universality tests. We perform global fits to the NP couplings in both one-dimensional (1D) and two-dimensional (2D) scenarios, where in the 1D case only a single WC deviates from its SM value, while in the 2D case two independent WCs are varied simultaneously. We then investigate how the resulting couplings affect the decay observables of $B_c \to D_s^{(*)}\mu^+\mu^-$. By examining how these fitted couplings influence $B_c \to D_s^{(*)}\mu^+\mu^-$ observables, we demonstrate that these decays provide complementary tests and can act as cross-checks of NP interpretations obtained from $b \to s \mu^+ \mu^-$ fits.

The remainder of the paper is organized as follows. In Section~II, we review the effective Hamiltonian describing semileptonic $b \to s \ell^+\ell^-$ transitions. Section~III presents the global fits to the NP Wilson coefficients in both 1D and 2D scenarios. The implications of the fitted couplings for $B_c\to D_s\mu^+\mu^-$ and $B_c\to D_s^*\mu^+\mu^-$ are discussed in Sections~IV and V, respectively. Finally, we summarize our results and conclude in Section~VI.  

\section{Effective Hamiltonian}
The general effective Hamiltonian for  rare semileptonic $b\rightarrow s \mu^+ \mu^-$ transition is given  as \cite{Ali:1999mm,Altmannshofer:2008dz,Sahoo:2015qha},
\begin{align}
\centering
   \mathcal{H}_{eff}&=-\frac{4\,G_F}{\sqrt{2}}V_{tb}\,V^*_{ts}\Big[C^{eff}_7\,\mathcal{O}_7+C_7^{'}\,\mathcal{O}_7^{'} \nn\\ &+\sum_{i=9,10,P,S}\big((C_i+C_i^{NP})\,\mathcal{O}_i+C_i^{'NP}\,\mathcal{O}_i^{'}\big)\Big],
\end{align}
where  $G_F$ is the Fermi coupling constant, $V_{ij}$ represents the Cabibbo-Kobayashi-Maskawa (CKM) matrix elements. Here, the four fermion operators $\mathcal{O}_i^{(\prime)}$, where $i=7,9,10,S,P$, are given as  
\begin{eqnarray*}
     \mathcal{O}_7^{(')}&&=\frac{e}{16\,\pi^2}\,m_b\,(\bar{s}\sigma_{\mu\,\nu}P_{R(L)}\,b)\,F^{\mu\,\nu},\\
     \mathcal{O}_9^{(')}&&=\frac{e^2}{16\,\pi^2}\,(\bar{s}\gamma_{\mu}P_{L(R)}\,b)\,(\bar{\mu}\gamma^{\mu}\mu),\\
      \mathcal{O}_{10}^{(')}&&=\frac{e^2}{16\,\pi^2}\,(\bar{s}\gamma_{\mu}P_{L(R)}\,b)\,(\bar{\mu}\gamma^{\mu}\,\gamma_5\,\mu),\\
      \mathcal{O}_S^{(')}&&=\frac{e^2}{16\,\pi^2}\,m_b\,(\bar{s}P_{R(L)}\,b)\,(\bar{\mu}\mu),\\
     \mathcal{O}_P^{(')}&&=\frac{e^2}{16\,\pi^2}\,m_b\,(\bar{s}P_{R(L)}\,b)\,(\bar{\mu}\gamma_5\mu),
\end{eqnarray*}
where $P_{R,L}=(1\pm \gamma_5)/2$ are the chirality operators and $C_i^{(\prime)}$'s are the corresponding Wilson coefficients\cite{PhysRevD.61.074024}. The SM does not include contributions from primed Wilson coefficients or (pseudo)scalar coefficients. These coefficients only appear in theories beyond the SM.

\section{Fitting new couplings to $b \to s \mu^+ \mu^-$ transitions}
Assuming the presence of new physics in the muon sector to address the observed discrepancies in $b \to s \mu^+ \mu^-$ transitions, we begin our analysis with a model-independent approach. We focus on the associated observables that exhibit deviations from the SM predictions. In particular, we consider scenarios where new physics effects may enter through the vector Wilson coefficients $C_9^{(\prime)\,\mathrm{NP}}$ and the axial--vector coefficients $C_{10}^{(\prime)\,\mathrm{NP}}$ which drive the $b \to s \mu^+ \mu^-$ decay observables. Contributions from the electromagnetic dipole operator $O_{7}^{(')}$, while relevant for radiative processes such as $b \to s \gamma$, play only a subleading role for the observables of interest and are therefore not included in our fit. The global fit is thus performed by varying the Wilson coefficients $C_{9}^{NP}$ and $C_{10}^{NP}$ (together with their possible primed counterparts), while keeping $C_{7}^{(')}$ fixed to its SM value. The set of \( b \to s \mu^+ \mu^- \) observables included in our fit are outlined below.
\begin{itemize}
    \item \textbf{Flavor specific $b \to s \mu^+ \mu^-$ observables}
    
\ding{43} \textit{Branching ratios}:
    
We include in our analysis the updated branching ratio of $B_s \to \mu^+ \mu^-$, reported by the LHCb collaboration at Moriond 2021~\cite{LHCb:2021awg}. 
The measured value, \(\mathcal{B}(B_s \to \mu^+ \mu^-) =(3.09^{+0.46+0.15}_{-0.43-0.11}) \times 10^{-9}\), 
is in close agreement with the previous world average~\cite{ATLAS2020}, which was obtained by combining results from the ATLAS~\cite{ATLAS:2018cur}, CMS~\cite{CMS:2019bbr}, and LHCb~\cite{LHCb:2017rmj} collaborations. 
In addition, we consider the branching ratios for $B \to K^{(*)}\mu^+\mu^-$~\cite{LHCb:2014cxe, LHCb:2016ykl} and $B_s \to \phi \mu^+ \mu^-$~\cite{LHCb:2021zwz}, evaluated in different \(q^2\) bins. 
Beyond the $b$-mesonic channels, we also include the baryonic decay $\Lambda_b \to \Lambda \mu^+ \mu^-$~\cite{LHCb:2015tgy}, mediated by the $b \to s \mu^+ \mu^-$ transition and studied across various \(q^2\) regions.

\ding{43} \textit{Angular observables in various $q^2$ bins}:

Here, we include the LHCb measurements~\cite{LHCb:2015svh, LHCb:2020gog, LHCb:2020lmf} of the forward--backward asymmetry $A_{FB}$, the longitudinal polarization fraction $F_L$, and the form factor independent observables $P_1$--$P_8^{\prime}$ in the $B^{0(+)} \to K^{*0(+)} \mu^+ \mu^-$ channels. 
We also incorporate the measurement of $F_L$ in the $B_s \to \phi \mu^+ \mu^-$ process~\cite{LHCb:2021xxq}. Also, the longitudinal and transverse contributions to the forward--backward asymmetry $A_{FB}$ in $\Lambda_b \to \Lambda \mu^+ \mu^-$ decays~\cite{LHCb:2018jna}, reported for the kinematic region $q^2 \in [15,\,20]~\text{GeV}^2$, are taken into account.

 \item \textbf{LFU sensitive observables} 
 
 We include the recent LHCb results on the isospin-partner ratios $R_{K_{S}^{0}}$ and $R_{K^{*+}}$~\cite{LHCb:2021lvy}, along with the measurements of $R_{K^{(*)}}$~\cite{LHCb:2022vje}, as well as the Belle measurements of the optimized angular observables $\Delta P'_{4}$ and $\Delta P'_{5}$~\cite{Belle:2016fev}.

\end{itemize}
Many analyses have been devoted to the global fit of the $b \to s \ell^+ \ell^-$ transitions~\cite{Descotes-Genon:2015uva,Alguero:2019ptt,Alguero:2021anc,Hurth:2023jwr,Wen:2023pfq,Alguero:2023jeh}, employing a model-independent effective field theory framework. Many of these works consider both 1D and 2D scenarios, in which the new physics explanations were found in the 1D hypotheses $C_9^{\rm NP}$, $C_9^{\rm NP}=-C_{10}^{\rm NP}$, and $C_9^{\rm NP}=-C_9^{\rm NP'}$, as well as in eight distinct 2D scenarios using the ``All'' and ``LFUV'' fits, while other possible scenarios were not considered. In this global analysis, we aim to explore all possible scenarios in both 1D and 2D hypotheses. 

In addition to the mesonic $b\to s\mu^+\mu^-$ observables, we also include the available $\Lambda_b\to\Lambda\mu^+\mu^-$ measurements in our global fit. Unlike Refs.~\cite{Alguero:2021anc, Alguero:2023jeh}, we account for the uncertainty associated with the $\Lambda_b$ production fraction, which affects the absolute normalization of the $\Lambda_b\to\Lambda\mu^+\mu^-$ branching ratio~\cite{London:2021lfn, Blake:2019guk}.

We note that the recent measurements of $R_{K}$ and $R_{K^{*}}$ are considerably closer to the SM expectations than earlier results and therefore no longer provide strong evidence for lepton flavor non-universality. Nevertheless, global analyses of the complete $b\to s\ell^{+}\ell^{-}$ dataset continue to show a preference for NP contributions in the muonic Wilson coefficients, driven mainly by branching ratio and angular observable measurements. In the present work, we therefore adopt the commonly used benchmark scenario in which NP affects only the muon sector and use the corresponding best-fit regions to investigate their implications for $B_c\to D_s^{(*)}\mu^{+}\mu^{-}$ observables.

\begin{figure*}[htbp]
\centering
\begin{subfigure}{0.328\textwidth}
\includegraphics[width=\linewidth]{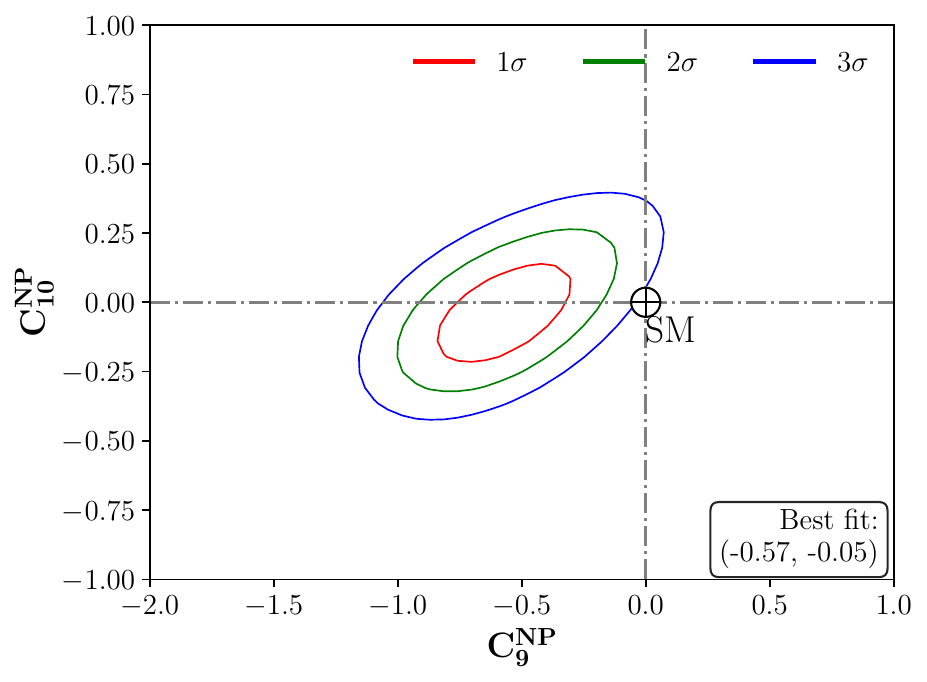}
\caption{$C_9^{NP}$ vs $C_{10}^{NP}$}
\end{subfigure}
\begin{subfigure}{0.328\textwidth}
\includegraphics[width=\linewidth]{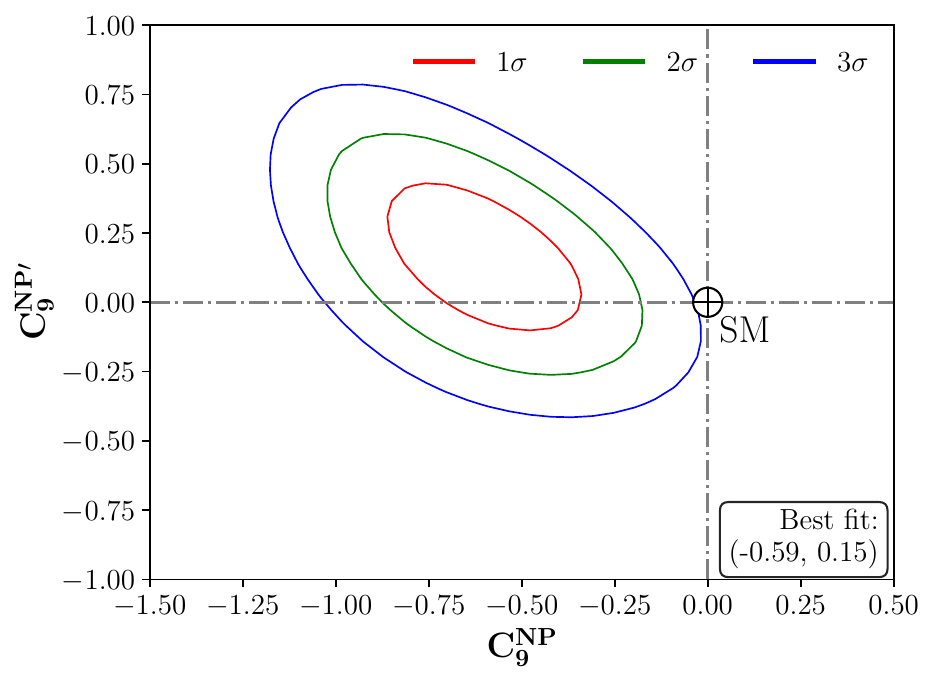}
\caption{$C_9^{NP}$ vs $C_{9}^{NP'}$}
\end{subfigure}
\begin{subfigure}{0.328\textwidth}
\includegraphics[width=\linewidth]{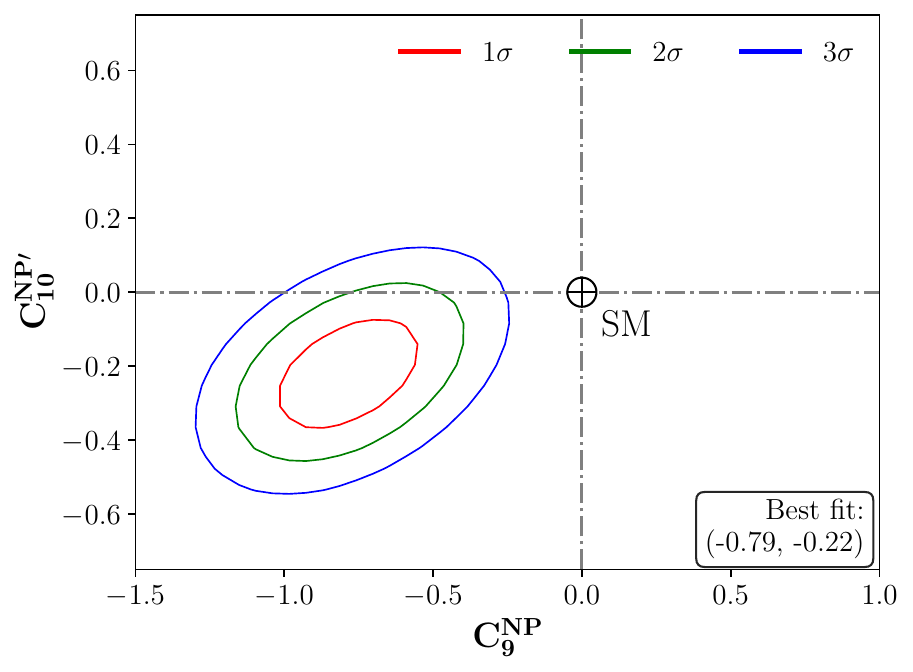}
\caption{$C_9^{NP}$ vs $C_{10}^{NP'}$}
\end{subfigure}

\begin{subfigure}{0.328\textwidth}
\includegraphics[width=\linewidth]{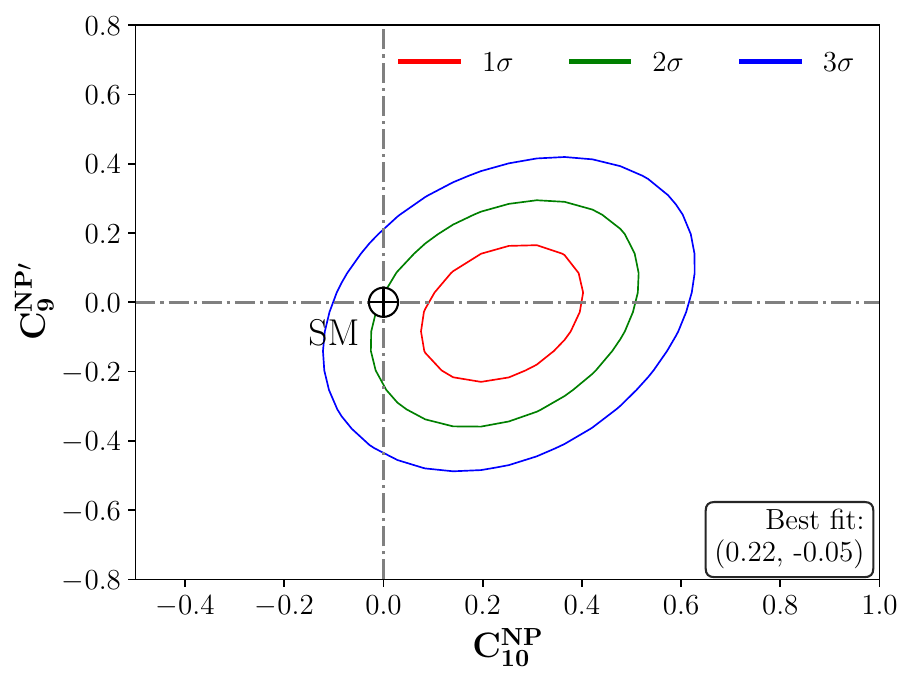}
\caption{$C_{10}^{NP}$ vs $C_{9}^{NP'}$}
\end{subfigure}
\begin{subfigure}{0.328\textwidth}
\includegraphics[width=\linewidth]{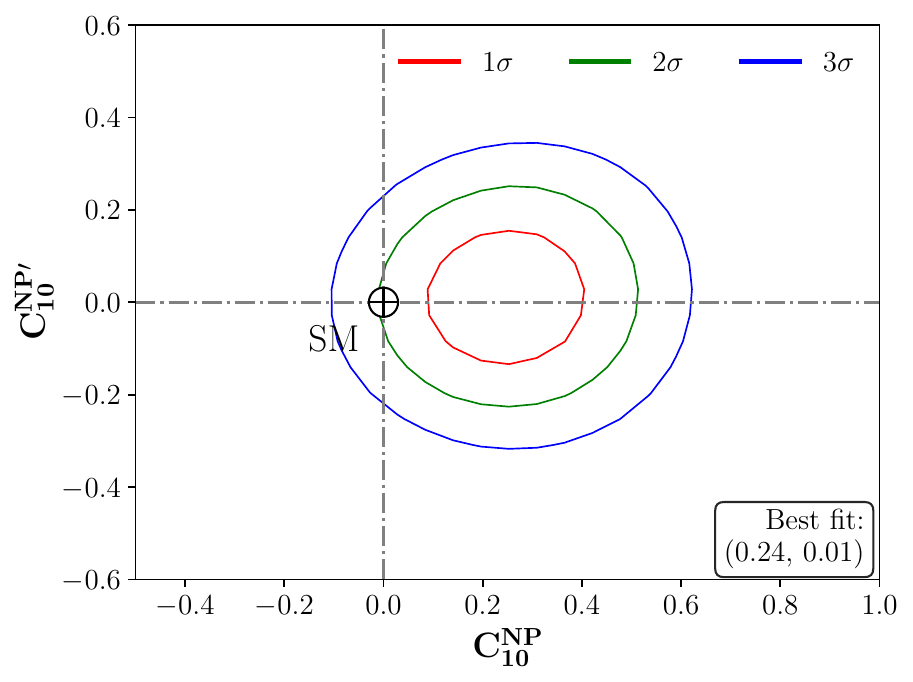}
\caption{$C_{10}^{NP}$ vs $C_{10}^{NP'}$}
\end{subfigure}
\begin{subfigure}{0.328\textwidth}
\includegraphics[width=\linewidth]{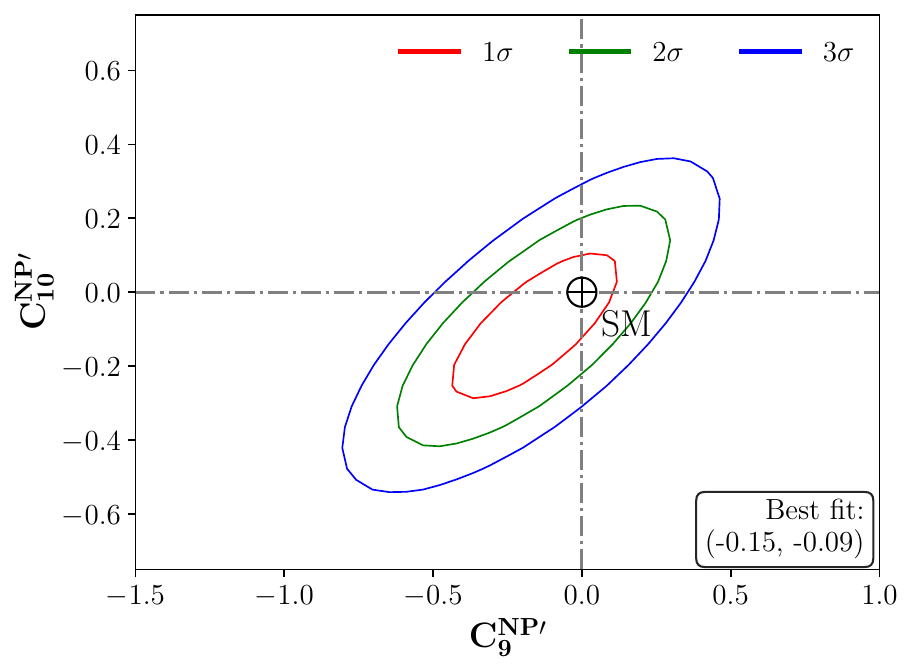}
\caption{$C_{9}^{NP'}$ vs $C_{10}^{NP'}$}
\end{subfigure}
\begin{subfigure}{0.328\textwidth}
\includegraphics[width=\linewidth]{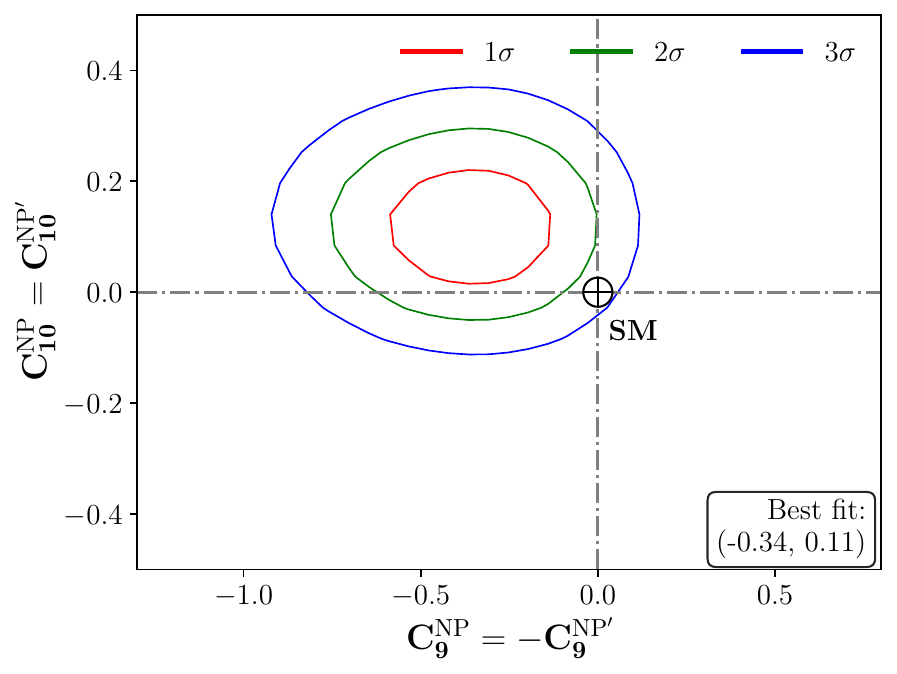}
\caption{$C_9^{NP}$= - $C_{9}^{NP'}$ vs $C_{10}^{NP}$= $C_{10}^{NP'}$}
\end{subfigure}
\begin{subfigure}{0.328\textwidth}
\includegraphics[width=\linewidth]{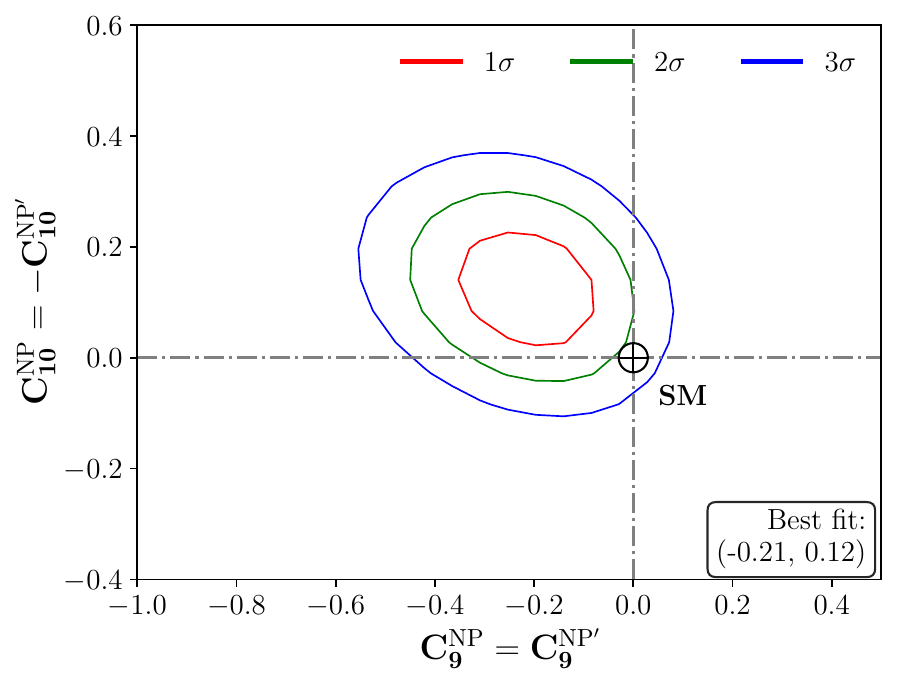}
\caption{$C_9^{NP}$ = $C_{9}^{NP'}$ vs $C_{10}^{NP}$ = - $C_{10}^{NP'}$}
\end{subfigure}
\begin{subfigure}{0.328\textwidth}
\includegraphics[width=\linewidth]{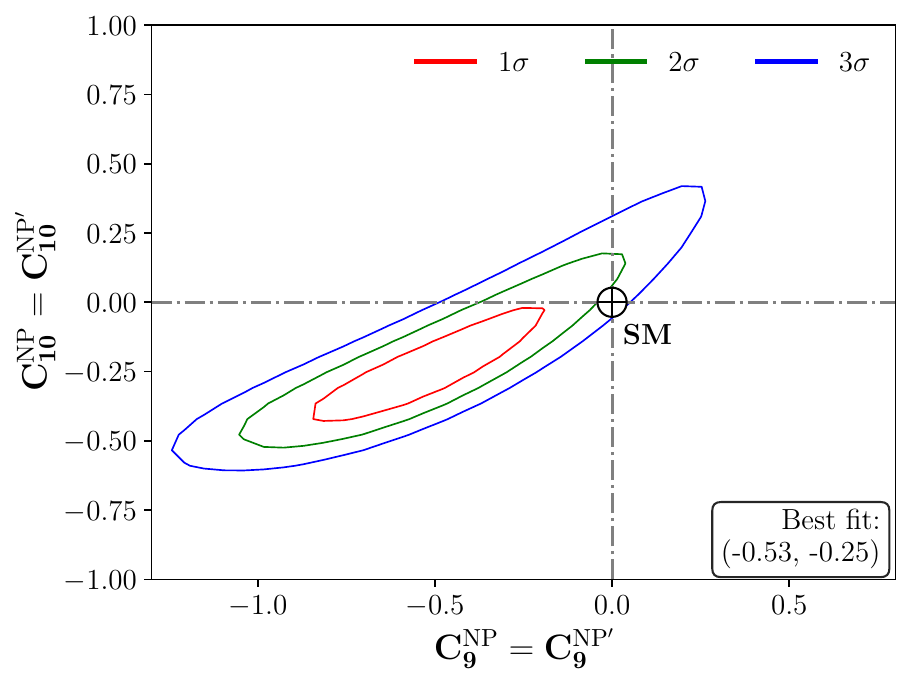}
\caption{$C_9^{NP}$= $C_{9}^{NP'}$ vs $C_{10}^{NP}$= $C_{10}^{NP'}$}
\end{subfigure}

\begin{subfigure}{0.328\textwidth}
\includegraphics[width=\linewidth]{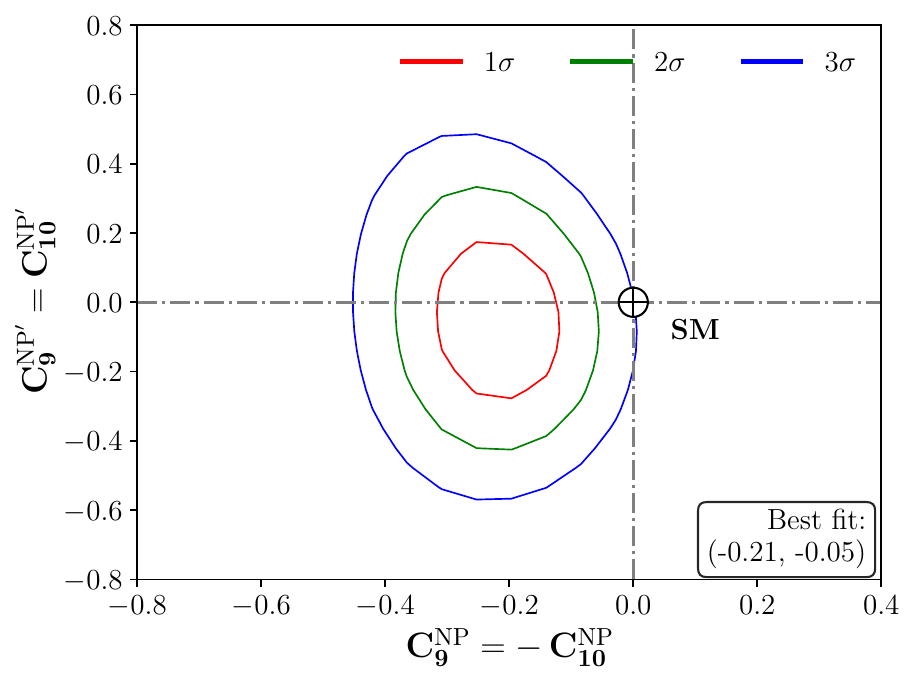}
\caption{$C_9^{NP}$= - $C_{10}^{NP}$ vs $C_{9}^{NP'}$= $C_{10}^{NP'}$}
\end{subfigure}
\begin{subfigure}{0.328\textwidth}
\includegraphics[width=\linewidth]{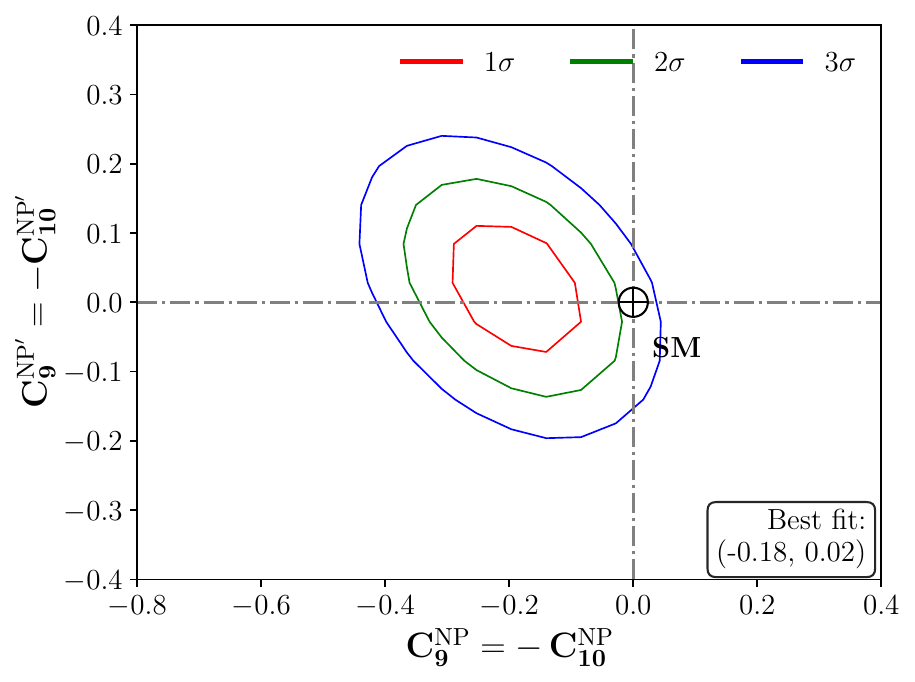}
\caption{$C_9^{NP}$ = - $C_{10}^{NP}$ vs $C_{9}^{NP'}$ = - $C_{10}^{NP'}$}
\end{subfigure}

\caption{Allowed parameter space for selected new physics scenarios in the $b \to s \mu^+\mu^-$ transitions.}
\label{fig:NP_contours_set2}
\end{figure*}

We now proceed with the analysis of new physics operators, specifically \(\mathcal{O}_9^{(\prime)NP}\) and \(C_{10}^{(\prime)NP}\), examining both one-dimensional (1D) and two-dimensional (2D) scenarios, where new physics is considered solely within the $\mu$-sector.  In this study, we employ all the observables discussed previously and conduct our analysis using the \texttt{flavio} package \cite{Straub:2018kue}. A complete analysis of the fit results derived from the $b \to s \mu \mu$ data is presented in Table~\ref{tab:1Dfits} for one-dimensional scenarios and Table~\ref{tab:2Dfits} for two-dimensional scenarios. The scenarios for all possible new physics couplings are labeled from S-I to S-XI. Additionally, we report the best-fit values along with their $1\sigma$ uncertainties, the pull = $\sqrt{\chi_{\rm SM}^2-\chi^2_{\rm best-fit}}$ values, and the p-values ($\%$). We constrain the new physics parameter space and illustrate the $1\sigma$, $2\sigma$, and $3\sigma$ allowed regions from the $b \to s \mu^+ \mu^-$ data, shown as red, blue, and green contours, respectively. The details are shown in Fig.~\ref{fig:NP_contours_set2}.
The 1D fits indicate that the scenarios with $C_9^{\rm NP}$ (S–I) and $C_9^{\rm NP}=-C_{10}^{\rm NP}$ (S–VI) are most preferred, with the larger pulls and highest p-values, showing strong sensitivity of the data to these coefficients. Primed coefficients and $C_{10}^{\rm NP}$ alone have smaller pulls, indicating a subdominant impact. The 2D fits indicate that the scenarios  $(C_9^{\rm NP},C_{10}^{\rm NP})$ (S--I), $(C_9^{\rm NP},C_{10}^{'\rm NP})$ (S--III) and $(C_9^{\rm NP}=-C_{10}^{\rm NP},C_{9}^{'\rm NP}=C_{10}^{'\rm NP})$ (S--X) are most favored, with the largest pulls and highest p-values, highlighting strong sensitivity to vector and axial-vector NP contributions. Scenarios involving only primed coefficients or $C_{10}^{\rm NP}$ with primed partners yield smaller pulls, indicating a subdominant effect. Overall, the data clearly favors NP contributions involving $C_9^{\rm NP}$, either alone or in combination, as the dominant source of deviations in $b \to s \mu^+ \mu^-$ transitions.\\
Now, our aim is to investigate the impact of NP on the observables of the decay process $B_c \to D_s^{(*)} \ell^+ \ell^-$. The analysis is carried out within an effective field theory framework that incorporates all possible vector and axial-vector interactions, leading to the introduction of the WCs $C_{9,10}^{NP}$ and $C_{9,10}^{\prime NP}$.

\begin{table*}[htbp]
\centering
\caption{ Best-fit $[1\sigma]$, pull and p-value($\%$) of different single Wilson coefficient NP scenarios in $b \to s \mu^+ \mu^-$ transitions.}\label{tab:1Dfits}
\begin{tabular}{@{}ccccccr@{}}
\toprule[1.2pt] 
Scenario & Coefficient & Best-fit value [$1\sigma$] & Ref.~\cite{Alguero:2023jeh}  & Pull &  p-value ($\%$) \\
\hline
\hline
 \midrule 
S - I & $C_9^{\rm NP}$ & $ -0.55$ $[\substack{-0.39 \\ -0.70}]$ & $ -0.66 [\substack{-0.52 \\ -0.82}]$  & 3.46 & 20.0\\ 
S - II & $C_{10}^{\rm NP}$ & $ 0.28 $ $[\substack{0.38 \\ 0.18}]$& --& 2.32 & 13.0 \\
S - III & $C_{9}^{'\rm NP}$ & $ -0.06 $ $[\substack{0.06 \\-0.18}]$& --&   1.14 & 09.0\\ 
S - IV & $C_{10}^{\prime\,\rm NP}$ & $-0.04\,\bigl[\substack{+0.04 \\ -0.12}\bigr]$ & -- & 1 & 08.0 \\
S - V & $C_9^{\rm NP}=C_{10}^{\rm NP}$ &  $ -0.06 $  $[\substack{0.06 \\ -0.19}]$& --&   0.90 & 07.0  \\
S - VI & $C_9^{\rm NP}=-C_{10}^{\rm NP}$ &   $ -0.20 $  $[\substack{-0.13 \\ -0.26}]$ & $-0.19 $  $[\substack{-0.13 \\ -0.25}]$&   2.86 & 15.0 \\
S - VII & $C_{9}^{'\rm NP}=C_{10}^{'\rm NP}$ &   $ -0.08 $ $[\substack{0.05 \\ -0.21}]$& --&   0.50 & 06.0  \\
S - VIII & $C_{9}^{'\rm NP}=-C_{10}^{'\rm NP}$ &  $ -0.02 $ $[\substack{0.03 \\ -0.08}]$& --&   0.35 & 05.0 \\
S - IX & $C_9^{\rm NP}=-C_{9}^{'\rm NP}$ & $ -0.37$ $[\substack{-0.22 \\ -0.53}]$& $ -0.67 $ $[\substack{-0.47 \\ -0.87}]$&   2.36 & 13.0 \\
S - X & $C_9^{\rm NP}=-C_{10}^{\rm NP}=-C_{9}^{'\rm NP}=-C_{10}^{'\rm NP}$   & $ -0.17 $ $[\substack{-0.11 \\ -0.24}]$& --     & 2.57 & 14.0 \\
S - XI & $C_9^{\rm NP}=-C_{10}^{\rm NP}=C_{9}^{'\rm NP}=-C_{10}^{'\rm NP}$  & $-0.07$ $[\substack{-0.03 \\ -0.11}]$& --&   1.77  & 11.0 \\
\bottomrule[1.2pt] 
\end{tabular}
\end{table*}

\begin{table*}[htbp]
\centering
\caption{Best-fit values $[1\sigma]$, pull, p-value($\%$) for different combinations of two Wilson coefficient NP scenarios in $b \to s \mu^+ \mu^-$ transitions.}\label{tab:2Dfits}
\begin{tabular}{@{}cccccr@{}}
\toprule[1.2pt] 
Scenario & Coefficient & Best fit value [$1\sigma$] & Ref.~\cite{Alguero:2023jeh} & Pull &  p-value ($\%$)  \\ 
\hline
\hline
 \midrule 
 S - I & $(C_9^{\rm NP},C_{10}^{\rm NP})$ & $(-0.57 [\substack{-0.38 \\ -0.76}], -0.05[\substack{0.07 \\ -0.17}])$ & $(-0.78, -0.13)$ & 3.69& 18.0  \\ 
S--II & $(C_9^{\rm NP}, C_9^{\prime\,\rm NP})$ & $(-0.59 [\substack{-0.44 \\ -0.75}],\; 0.15 [\substack{ 0.20 \\ 0.10 }])$ &
$(-0.85, 0.38)$ &4.20 & 21.0 \\
S - III &  $(C_9^{\rm NP},C_{10}^{'\rm NP})$ &  ($-0.79[\substack{-0.64 \\ -0.94}]$, $-0.22[\substack{-0.12 \\ -0.32}]$ )  & (-0.78, -0.16) & 4.98 & 26.0\\ 
S - IV & $(C_{10}^{\rm NP},C_{9}^{'\rm NP})$ & ($0.22[\substack{0.33 \\ 0.11}] $, $ -0.05 [\substack{0.07 \\ -0.17}]$) & -- & 3.26 & 15.0\\ 
S - V & $(C_{10}^{\rm NP},C_{10}^{'\rm NP})$ & ($0.24[\substack{0.34 \\ 0.14}] $, $ 0.01 [\substack{0.10 \\ -0.08}]$)   & -- & 2.55 & 10.0 \\ 
S - VI & $(C_{9}^{'\rm NP},C_{10}^{'\rm NP})$ &  ($-0.15[\substack{0.04 \\ -0.34}] $, $ -0.09 [\substack{0.04 \\-0.22}]$)  & -- & 1.01 & 07.0 \\ 
S - VII & $(C_{9}^{\rm NP}=-C_{9}^{'\rm NP},C_{10}^{\rm NP}=C_{10}^{'\rm NP})$ & ($-0.34[\substack{-0.18 \\ -0.50}] $, $ 0.11 [\substack{0.18 \\ 0.04}]$) & (-0.69, 0.12) & 3.26 & 14.0\\ 
S - VIII & $(C_{9}^{\rm NP}=C_{9}^{'\rm NP}, C_{10}^{\rm NP}=-C_{10}^{'\rm NP})$ &  ($-0.21[\substack{-0.11 \\ -0.31}] $, $ 0.12 [\substack{0.19 \\ 0.05}]$) & -- & 3.96 & 19.0 \\ 
S - IX & $(C_{9}^{\rm NP}=C_{9}^{'\rm NP},C_{10}^{\rm NP}=C_{10}^{'\rm NP})$ &  ($-0.53[\substack{-0.33 \\ -0.73}] $, $ -0.25 [\substack{-0.12 \\ -0.38}]$) & -- & 2.78 & 12.0\\ 
S - X & $(C_{9}^{\rm NP}=-C_{10}^{\rm NP},C_{9}^{'\rm NP}=C_{10}^{'\rm NP})$ & ($-0.21[\substack{-0.15 \\ -0.27}] $, $ -0.05 [\substack{0.09 \\ -0.19}]$) & (-0.20, 0.16) & 3.01 & 13.0\\ 
S - XI & $(C_{9}^{\rm NP}=-C_{10}^{\rm NP},C_{9}^{'\rm NP}=-C_{10}^{'\rm NP})$ &  $-0.18[\substack{-0.11 \\-0.25}] $, $ 0.02 [\substack{0.08 \\ -0.04}]$)  & (-0.19, 0.01) & 3.60 & 17.0\\ 

\bottomrule[1.6pt] 
\end{tabular}
\end{table*}

\section{Analysis of $B_c\rightarrow D_s\mu^+\mu^-$ decay observables}
\begin{figure*}[htbp]
\centering
\hspace{-1.7cm}\includegraphics[width=18cm,height=7cm]{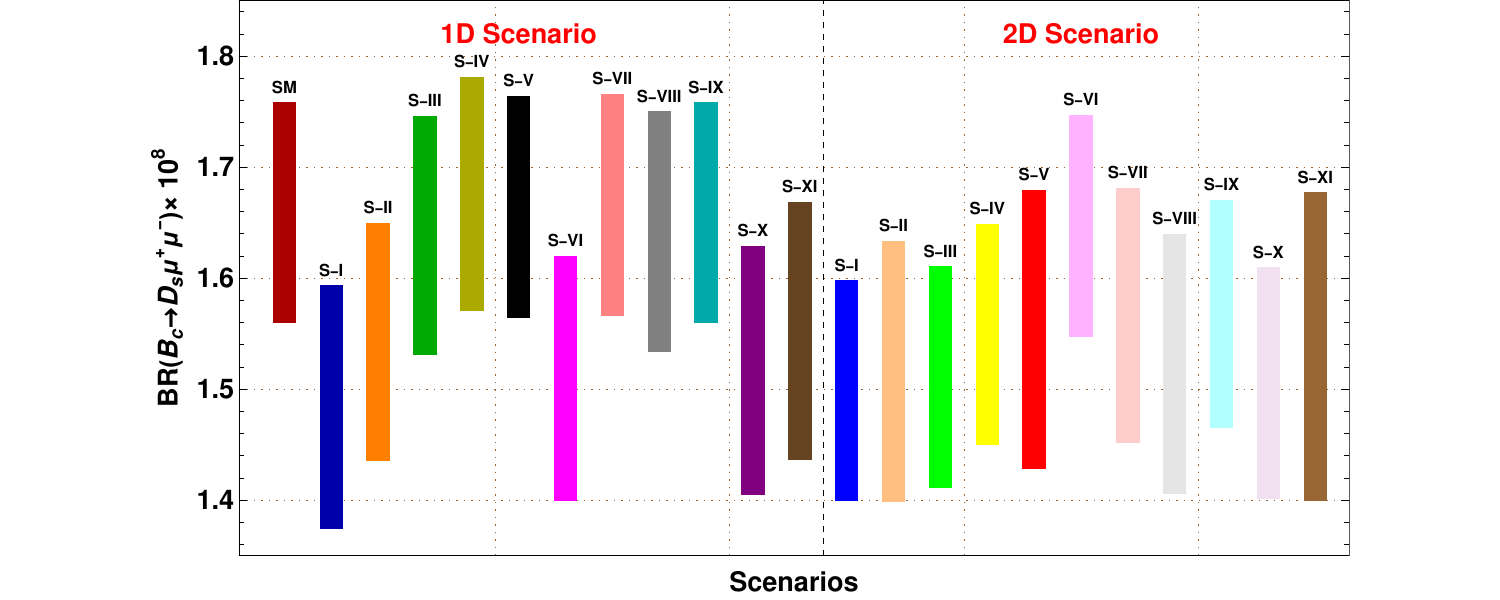}\\
\hspace{-1.7cm}\includegraphics[width=18cm,height=7cm]{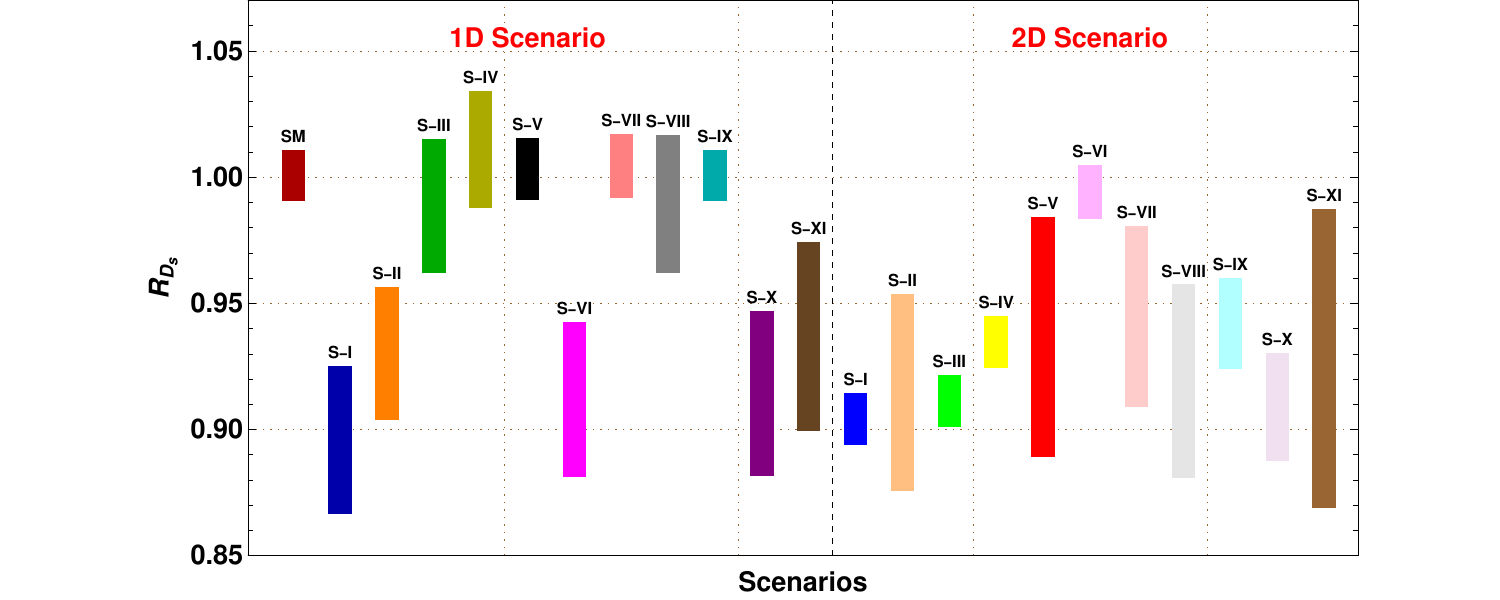}
\caption{Predicted range $(1\sigma)$ of the branching ratio (top) and lepton non-universality ratio $R_{D_s}$ (bottom) for the   $B_c \to D_s \mu^+ \mu^-$ process in the $q^2 \in [1,6]$ interval for all the new physics scenarios S-I to S-XI. Left: 1D scenario; Right: 2D scenario. {\color{blue}The SM prediction bands represent the theoretical uncertainties, while the NP bands include both the theoretical uncertainties and the uncertainties arising from the allowed $1\sigma$ ranges of the NP Wilson coefficients.}}
\label{Fig:BcDs BR}
\end{figure*}

After establishing the best-fit values and $1\sigma$ uncertainties for the new (axial)vector coefficients, we further analyze their impact on the branching ratio and angular observables of the  $B_c\rightarrow D_s\mu^+\mu^-$ decay mode.  This investigation seeks to reveal how these coefficients influence the decay dynamics, offering a deeper insight into the potential contributions of NP to this process. The $q^2$ differential decay rate for $B_c\rightarrow D_s\mu^+\mu^-$ process in the presence of new physics is given as \cite{Bobeth:2007dw} 
\begin{equation}
    \frac{d\Gamma(B_c\rightarrow D_s\mu^+\mu^-)}{dq^2}=2\,a_l (q^2)+\frac{2}{3}\,c_l (q^2)\,,
\end{equation} where, $a_l(q^2)$ and $c_l(q^2)$ are defined as
\begin{eqnarray*}
\centering
    a_l (q^2)&=&N_{D_s}\,\Big[q^2\,|F_p|^2+\frac{\lambda(m_{B_c}^2,m_{D_s}^2,q^2)}{4}(|F_A|^2+|F_V|^2)\\ &+&4m^2_l\,m^2_{B_c}|F_A|^2+2\,m_l(m_{B_c}^2-m^2_{D_s}+q^2)\,Re(F_p\,F_A^*)\Big],\\
    c_l(q^2) &&=-N_{D_s}\frac{\lambda(m_{B_c}^2,m_{D_s}^2,q^2)\,\beta_l^2}{4}(|F_A|^2+|F_V|^2)\,,
\end{eqnarray*}
with
\begin{eqnarray*}
\centering
   N_{D_s}&&=\frac{G_F^2\,\alpha_e^2\,|V_{tb}\,V_{ts}^*|^2}{2^9\,\pi^5\,m^3_{B_c}}\,\beta_l\,\sqrt{\lambda(m_{B_c}^2,m_{D_s}^2,q^2)}\,,
\end{eqnarray*}
the kinematic component  $\lambda(a,b,c)=a^2+b^2+c^2-2(a b+b c+ c a)$ and the mass correction factor $\beta_l=\sqrt{1-\frac{4\,m_l^2}{q^2}}$. Here the $F_{P,V,A}$ functions are the $q^2$ dependency form factors.\\
Now, using the required input parameters from particle data group (PDG) \cite{ParticleDataGroup:2024cfk} and the relativistic quark model form factor for the $B_c \to D_s$ transition from ref.\cite{Ebert:2010dv}, we investigate the branching ratio and LNU parameter of $B_c \to D_s \mu^+ \mu^-$ process in the SM and in the presence of new vector and axial-vector coefficients. Based on the $1\sigma$ allowed ranges of new coefficients in various 1D and 2D scenarios  (S-I to S-XI), the predicted $1\sigma$ range of the branching ratios of $B_c \to D_s \mu^+ \mu^-$ decay mode in the $q^2\in [1,6]$ interval is presented in Fig. \ref{Fig:BcDs BR} (top). 
In this figure, the SM result is represented in maroon color, and the prediction band reflects theoretical uncertainties, primarily arising from the form factors and CKM matrix elements. Fig. \ref{Fig:BcDs BR} (bottom) depicts our predictions for the LNU observable $R_{D_s}$, derived from the $1\sigma$ ranges of different 1D and 2D scenarios for the new coefficients.  The NP uncertainty bands include both the uncertainties arising from the allowed $1\sigma$ ranges of the NP Wilson coefficients and the theoretical uncertainties associated with the input parameters.

In the case of 1D scenarios, the  $C_9^{\rm NP}=-C_{10}^{\rm NP}$ scenario is distinguishable from others by exhibiting the maximum deviation from the SM predictions. The scenarios with only $C_9^{\rm NP}$, only $C_{10}^{\rm NP}$,  $(C_{9}^{\rm NP}=-C_{10}^{\rm NP}=-C_{9}^{'\rm NP}=-C_{10}^{'\rm NP})$ and $(C_9^{\rm NP}=-C_{10}^{\rm NP}=C_{9}^{'\rm NP}=-C_{10}^{'\rm NP})$ also exhibit significant deviations from the SM results, whereas the predictions for the remaining scenarios exhibit minimal deviation or remain consistent with the SM. In the low $q^2\in[1,6]~{\rm GeV}^2$ region, $R_{D_s}$ ratio is predicted to be less than one for the scenarios S-I, S-II, S-VI, S-X, S-XI, while predictions from other scenarios are either almost equal to or slightly greater than one. We also perform a numerical comparison between our best-fit values and the previous results of Ref.~\cite{Alguero:2023jeh}. We find that the scenarios $C_9^{\rm NP}$ (S-I) and $C_9^{\rm NP}=-C_9^{\prime,\rm NP}$ (S-IX) show a reduction in the preferred NP magnitude compared with Ref.~\cite{Alguero:2023jeh}, while the $C_9^{\rm NP}=-C_{10}^{\rm NP}$ scenario (S-VI) exhibits excellent agreement with nearly identical best-fit values and uncertainties.\\
Analysis of 2D scenarios in the branching ratio indicates that the combinations of WCs $(C_9^{\rm NP},C_{10}^{\rm NP})$, $(C_{9}^{\rm NP}=C_{9}^{'\rm NP},C_{10}^{\rm NP}=-C_{10}^{'\rm NP})$ and $(C_{9}^{\rm NP}=-C_{10}^{\rm NP},C_{9}^{'\rm NP}=C_{10}^{'\rm NP})$ provide the most significant shifts from the SM results, whereas the scenario with $(C_{9}^{'\rm NP},C_{10}^{'\rm NP})$  yields minimal deviation. All the scenarios predict values of the $R_{D_s}$ parameter to be less than one with remarkable deviations in the presence of NP couplings. Similarly, for the two-dimensional fits, the best-fit values are generally reduced in magnitude compared with Ref.~\cite{Alguero:2023jeh}. However, the overall hierarchy and the relative preference among the favored NP scenarios remain unchanged.

\section{Inspection of $B_c\rightarrow D_s^{*}\mu^+\mu^-$ decay}

\begin{figure*}[htbp]
\centering
\hspace{-1.7cm}\includegraphics[width=18cm,height=7cm]{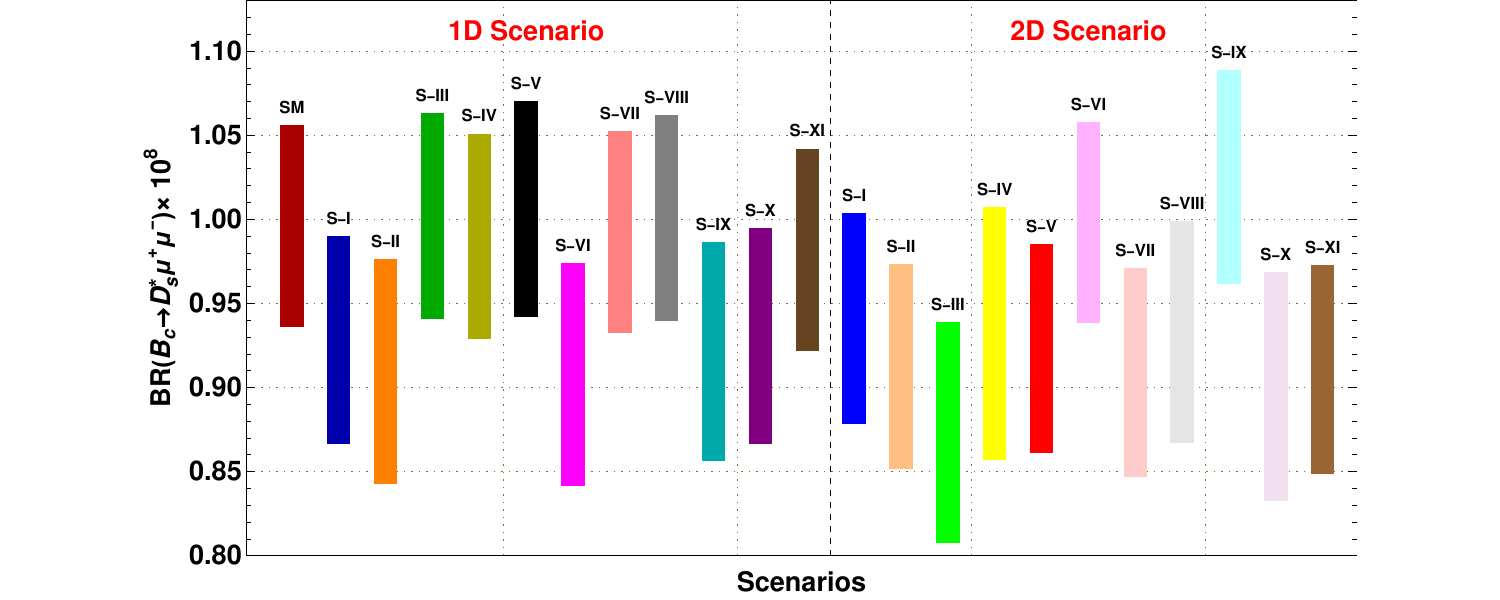} \\
\hspace{-1.7cm}\includegraphics[width=18cm,height=7cm]{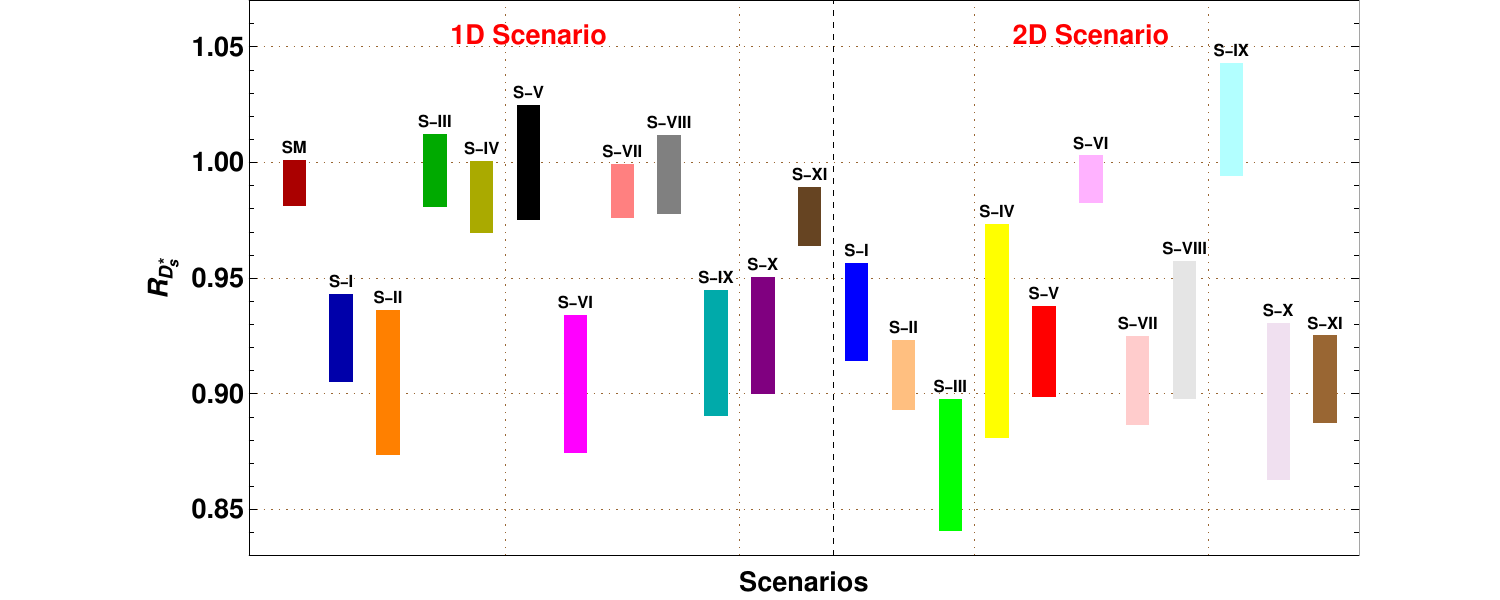}
\caption{Predicted 1$\sigma$ range of $BR(B_c \to {D_s^{(*)}} \mu^+\mu^-)$ (top) and lepton non-universality observable $R_{D_s^*}$ (bottom) across new physics scenarios S-I to S-XI. Left: 1D scenario; Right: 2D scenario. }
\label{Fig:Br Bc to Dsstar}
\end{figure*}

For $B_c\rightarrow D_s^{(*)}\mu^{+}\mu^{-}$ decay, the $q^2$-dependent differential decay width is given as\cite{Altmannshofer_009}
\begin{eqnarray*}
    \frac{d\Gamma}{dq^2}=\frac{1}{4}\Big[3\,I_1^c+6\,I_1^s-I_2^c-2\,I_2^s\Big],
\end{eqnarray*}
where $I_i^{c,s}$'s are  the angular coefficients \cite{Altmannshofer_009}\,. 
In addition to examining the branching ratios, we also investigate the following observables to analyze the composition of new physical entities.
\begin{itemize}
    \item The lepton forward-backward asymmetry:
\begin{equation*}
    A_{FB}(q^2)=\frac{3\,I_6}{3\,I_1^c+6\,I_1^s-I_2^c-2\,I_2^s}\,.
\end{equation*}

    \item The longitudinal and transverse polarization fractions of $D_s^*$:
    \begin{equation*}
    F_L(q^2)=\frac{3\,I_1^c-I_2^c}{3\,I_1^c+6\,I_1^s-I_2^c-2\,I_2^s},~~~~~ F_T(q^2)=1-F_L(q^2)\,.
\end{equation*}
\item The Lepton Flavor Universality Violation (LFUV) ratio:
      \begin{equation*}
    R_{D_s^*}(q^2)=\frac{d\Gamma(B_c\rightarrow D_s^{(*)}\mu^+\mu^-)/dq^2}{d\Gamma(B_c\rightarrow D_s^{(*)}\,e^+\,e^-)/dq^2}\,.
\end{equation*}
\item The form factor independent (FFI) observables \cite{Descotes_Genon_2013, Matias_2012}:
\begin{eqnarray*}
\centering
P_1&=&\frac{I_3}{2\,I_2^s},~~~~~~~~~~~ P_2=\beta_l\,\frac{I_6^s}{8\,I_2^s},~~~~~~~~~~~~ P_3=-\frac{I_9}{4\,I_2^s},\\
   P_4^{'} &=&\frac{I_4}{\sqrt{-I_2^c\,I_2^s}}, ~~~~P_5^{'}=\frac{I_5}{2\sqrt{-I_2^c\,I_2^s}}, 
    ~~~~~~~P_8^{'}=\frac{-I_8}{\sqrt{-I_2^c\,I_2^s}}.
\end{eqnarray*}
\end{itemize}

\begin{figure*}[htb]
\centering
\hspace{-1.7cm}\includegraphics[width=18cm,height=7cm]{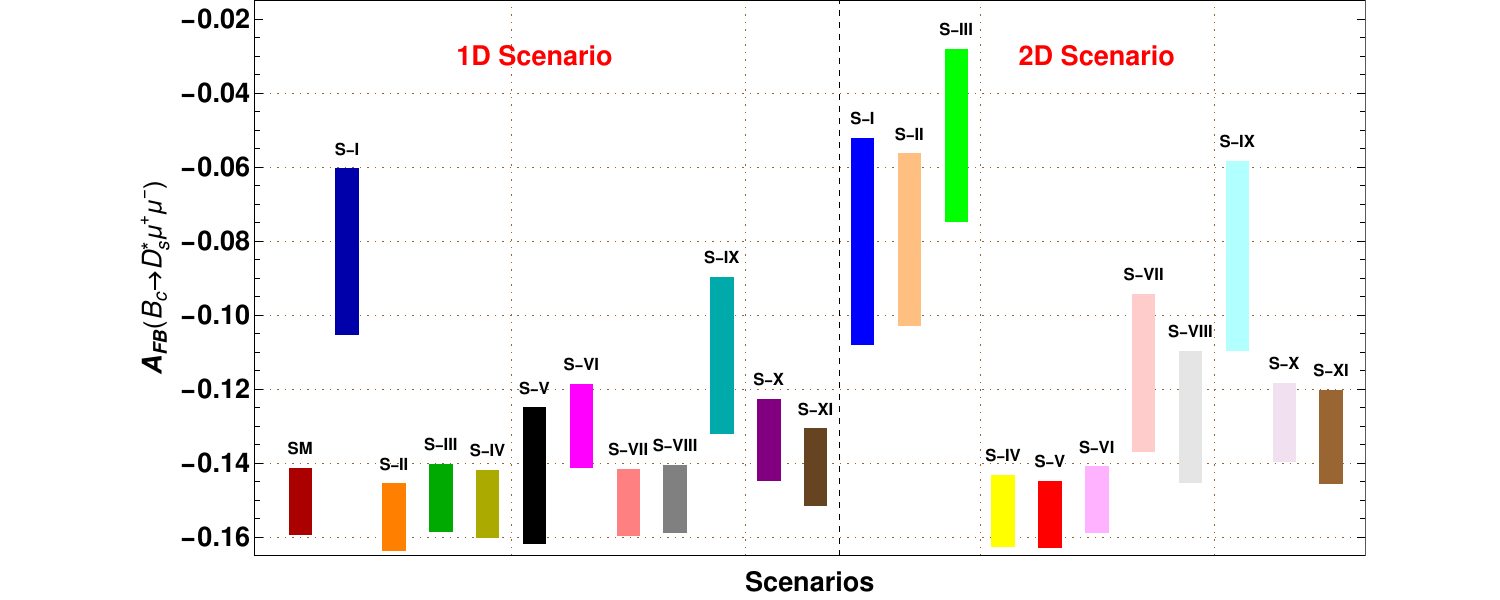} \\
\hspace{-1.7cm}\includegraphics[width=18cm,height=7cm]{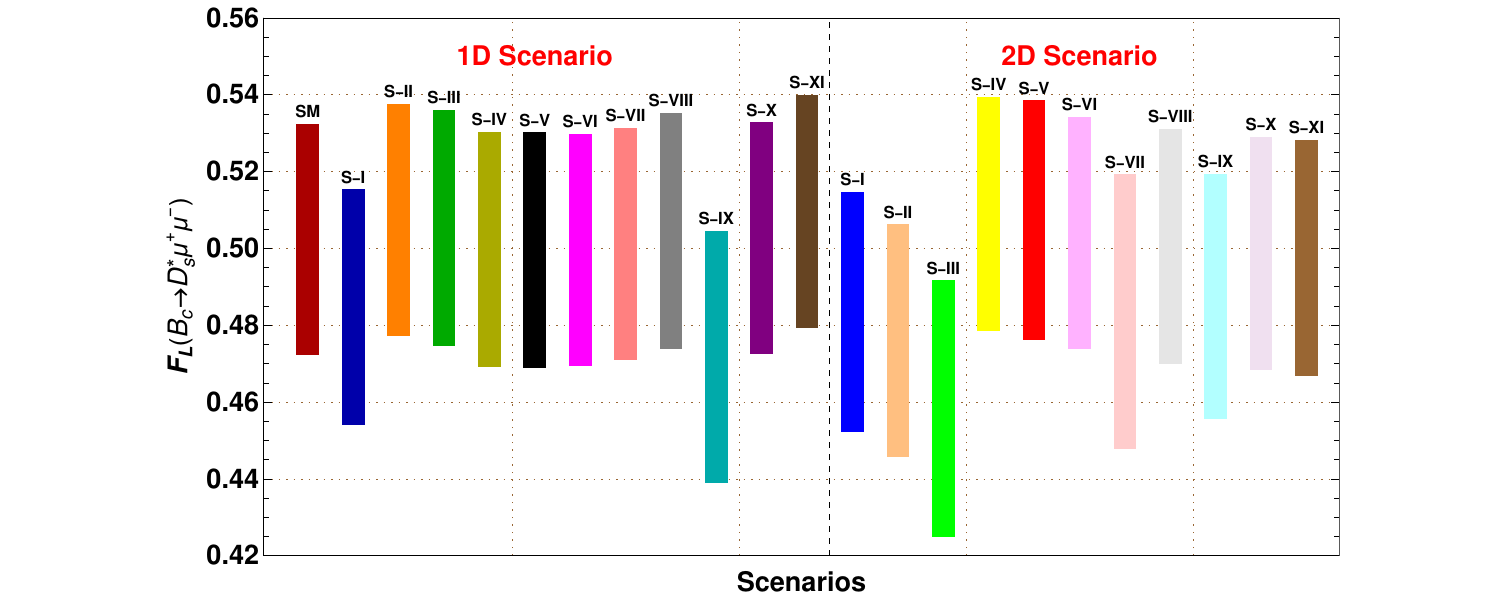}
\caption{Predicted range $(1\sigma)$ of forward-backward asymmetry (top) and longitudinal polarization fraction (bottom) of $B_c \to D_s^{(*)} \mu^+ \mu^-$ process for all the new physics scenarios. Left: 1D scenario; Right: 2D scenario.} \label{Fig:Afb}
\end{figure*}

To perform the numerical analysis of $B_c \to {D_s^{*}} \mu^+ \mu^-$ channel, we use the form factors  computed in the relativistic quark model from Ref. \cite{Ebert:2010dv} and all required input parameters, such as mass, lifetime, and CKM matrix elements, from PDG  \cite{ParticleDataGroup:2024cfk}. Using the $1\sigma$ range of new coefficients for both 1D and 2D scenarios, we graphically represent the predicted values for various quantities related to the $B_c \to {D_s^*} \mu^+ \mu^-$ mode, including the branching ratio, LNU parameter $R_{D_s^*}$  (Fig. \ref{Fig:Br Bc to Dsstar}), and the forward-backward asymmetry as well as the longitudinal polarization asymmetry (Fig. \ref{Fig:Afb}).

\begin{figure*}[htbp]
\centering
\hspace{-1.7cm}\includegraphics[width=18cm,height=7cm]{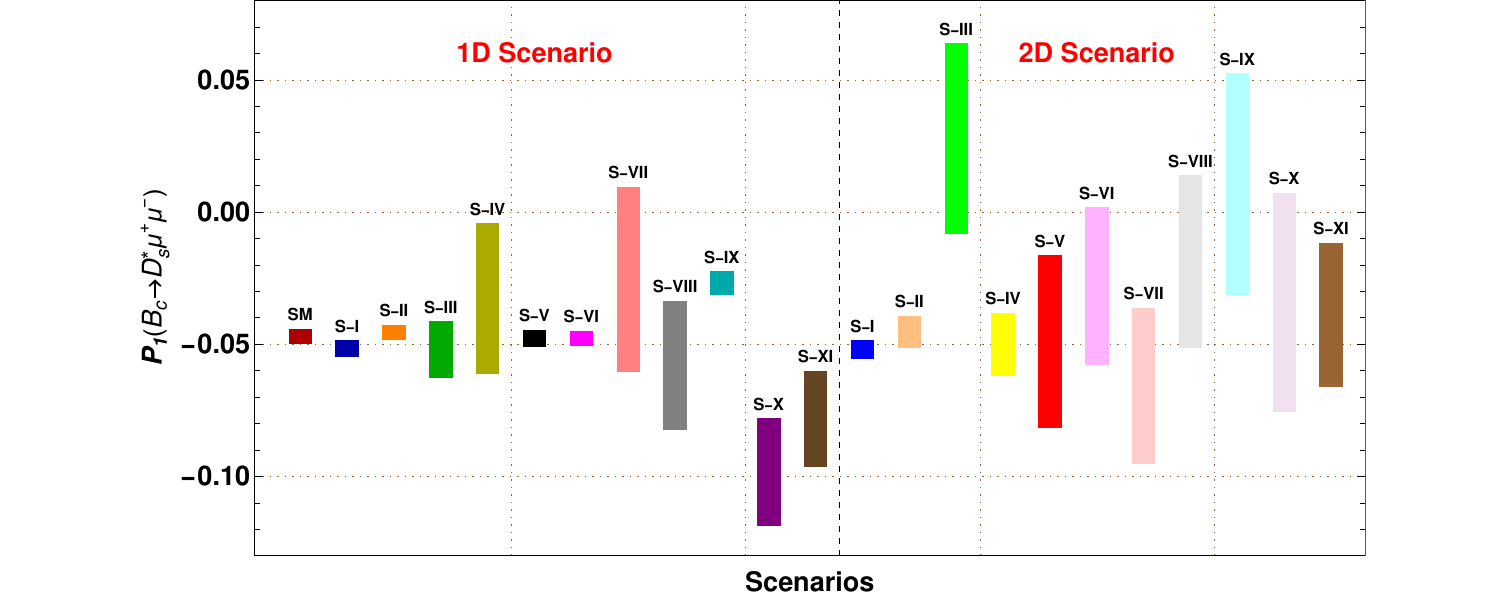}\\
\hspace{-1.7cm}\includegraphics[width=18cm,height=7cm]{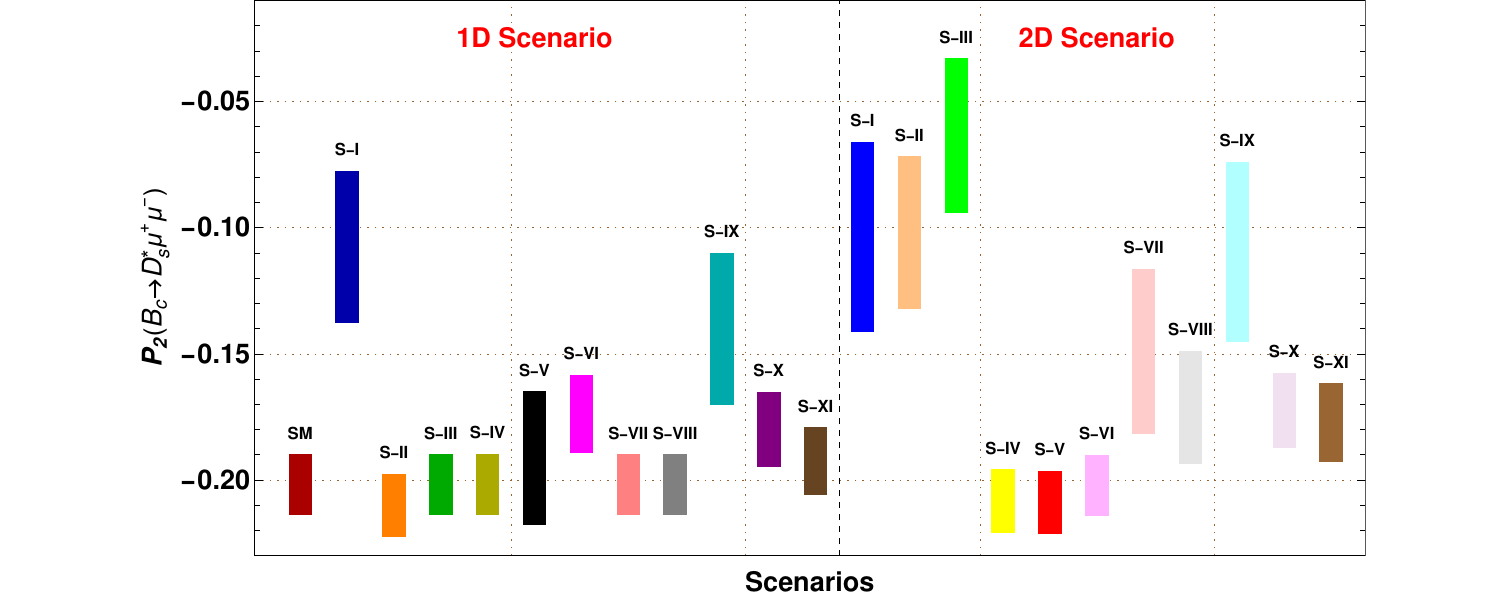} \\
\hspace{-1.7cm}\includegraphics[width=18cm,height=7cm]{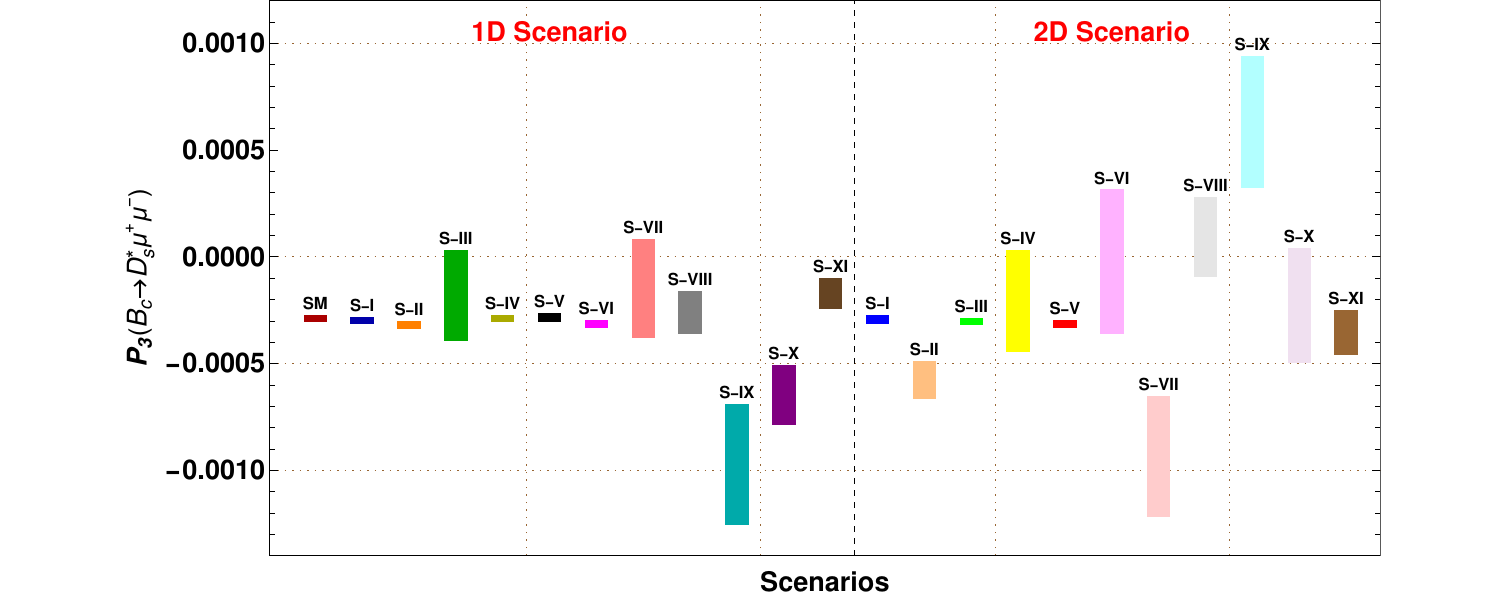}
\caption{Predicted range $(1\sigma)$ of  $P_1$ (top), $P_2$ (middle) and $P_3$ (bottom) observables for all the new physics scenarios.  Left: 1D scenario; Right: 2D scenario.}
\label{Fig:P1}
\end{figure*}

\begin{figure*}[htbp]
\centering
\hspace{-1.7cm}\includegraphics[width=18cm,height=7cm]{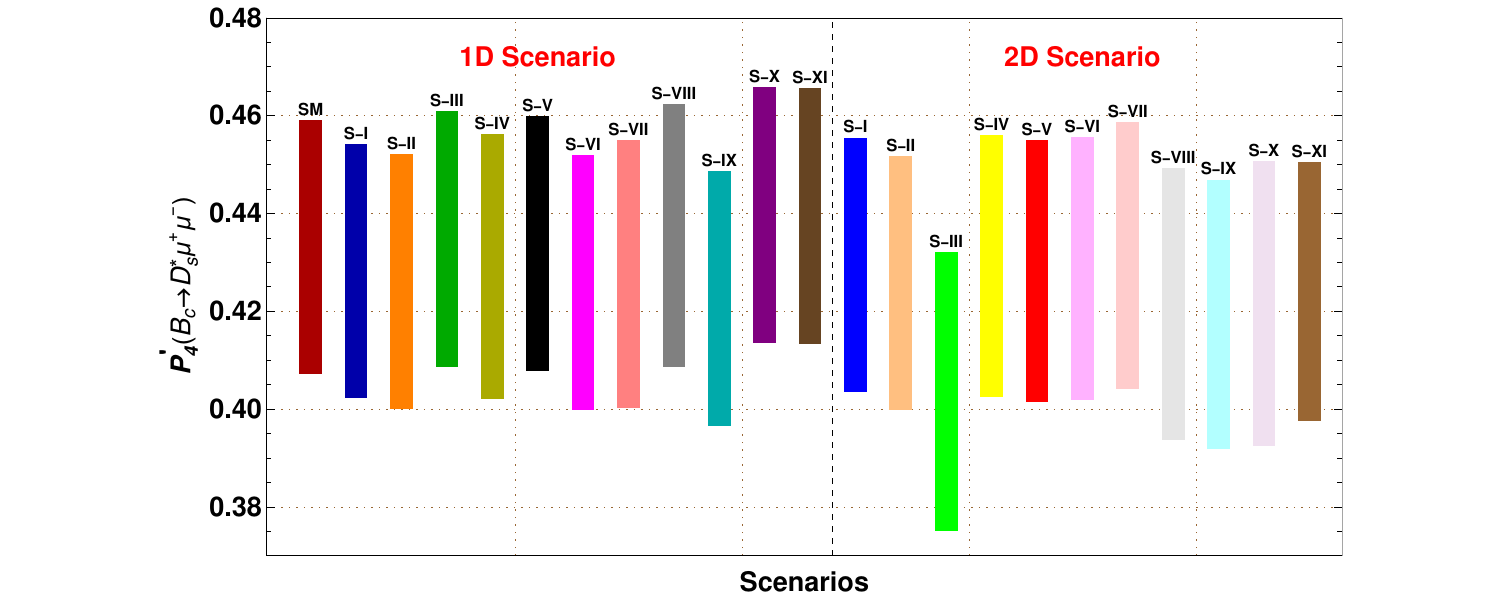}\\
\hspace{-1.7cm}\includegraphics[width=18cm,height=7cm]{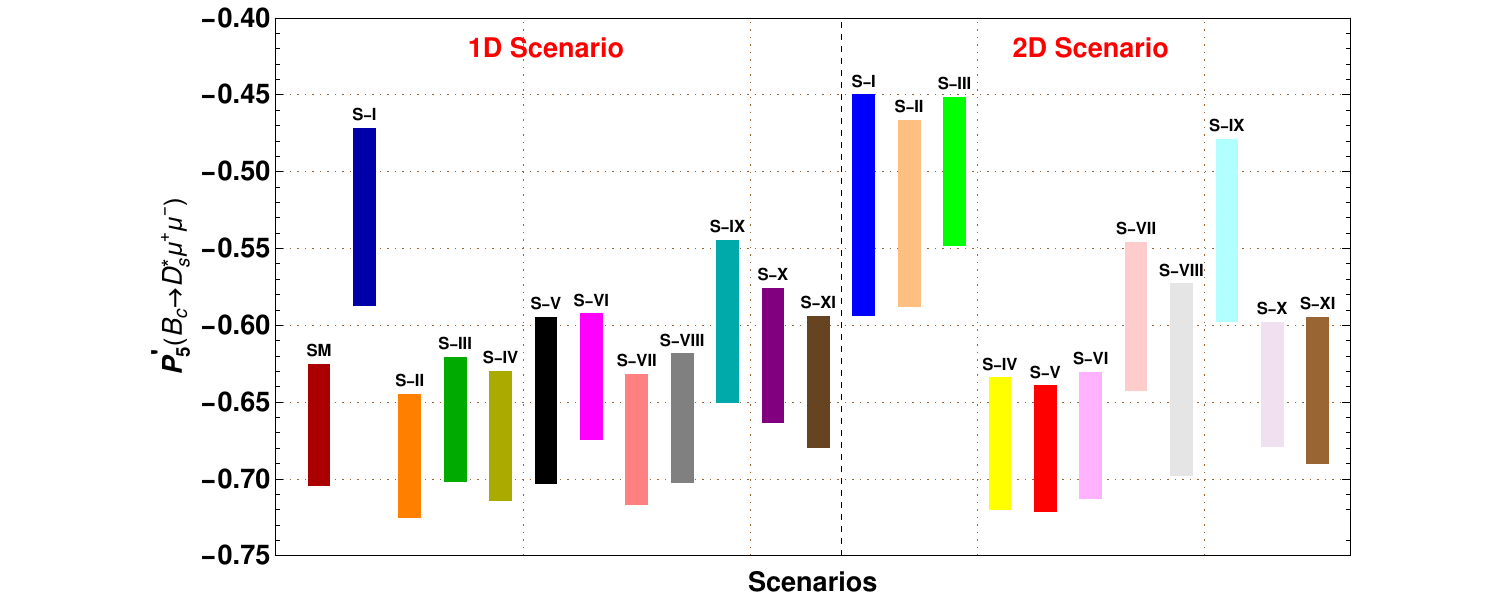} \\
\hspace{-1.7cm}\includegraphics[width=18cm,height=7cm]{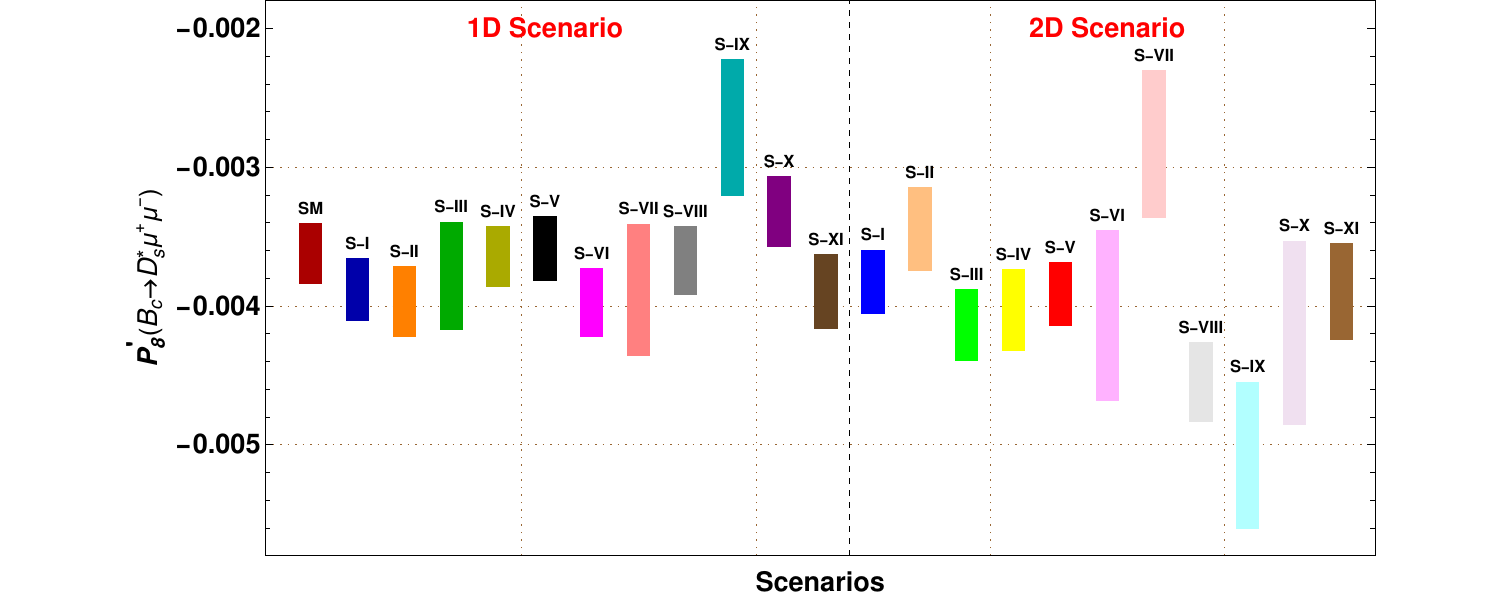}
\caption{Predicted range $(1\sigma)$ of $P_4^{\prime}$ (top), $P_5^{\prime}$ (middle) and $P_8^{\prime}$ (bottom) observables for all the new physics scenarios.\hspace{1.2cm} Left: 1D scenario; Right: 2D scenario.}
\label{Fig:P4p}
\end{figure*}

The $1\sigma$ predicted values for additional clean observables $P_1$, $P_2$ and $P_3$, which are independent of form factors, are graphically represented in Fig. \ref{Fig:P1}. Similarly, the observables $P_4^{'}$  $P_5^{'}$ and $P_8^{'}$ are shown in Fig. \ref{Fig:P4p}.
In these figures, the maroon color band stands for the SM prediction, which accounts for theoretical uncertainties,  while the other colored bands illustrate the $1\sigma$ range of predictions based on the new coefficients. 
Due to the marginal effects of the NP couplings on the $P_6^{'}$ observable, we did not include it in our analysis.

Within the frameworks of 1D scenarios:
\begin{itemize}
    \item The scenario S-VI, where $C_9^{\rm NP} = -C_{10}^{\rm NP}$ exhibits the maximum shift in the branching ratio for $B_c \to D_s^{*} \mu^+ \mu^-$  in the low $q^2$ bin, while other scenarios show only marginal deviations from the SM prediction. 
    \item All the scenarios predict $R_{D_s^*}<1$, suggesting the presence of lepton flavor universality violation in $b \to s \mu^+ \mu^-$ transition. Among these, the scenario with $C_9^{\rm NP} = -C_{10}^{\rm NP}$ shows the largest deviation from one, thereby supporting lepton non-universality.  
    \item For the forward-backward asymmetry observable, significant deviations from the SM are observed for scenarios with only $C_9^{\rm NP}$, $C_9^{\rm NP}=-C_{10}^{\rm NP}$ and $C_9^{\rm NP}=-C_{9}^{'\rm NP}$ coefficients. 
    \item The prediction for longitudinal polarization asymmetry with $C_9^{\rm NP}=-C_{9}^{'\rm NP}$ shows the comparatively large deviation from the SM result than other scenarios.
    \item In the analysis of form factor independent clean observables in the low $q^2$ bin, we observe that: \\(i) Scenarios S-X and S-XI are particularly sensitive to the $P_1$ observable, \\(ii) The $P_2$ observable shows maximum deviations for scenarios S-I, S-VI and S-IX,\\ (iii) The $P_3$ observable deviates most significantly for the S-IX scenario,\\ (iv) No such notable impact is observed in a notable impact on the $P_4^{'}$ observable, \\ (v) The $P_5^{'}$ observable receives the maximum contribution from scenarios S-I, S-VI and S-IX scenarios, \\ (vi) Scenarios S-VI and S-IX are crucial for the $P_8^{'}$ observable. 
\end{itemize}
In the realm of 2D scenarios:
\begin{itemize}
    \item The branching ratio of $B_c \to D_s^{*} \mu^+ \mu^-$ differs significantly for the new scenarios S-I, S-X and S-XI. 
    \item All scenarios predict that the $R_{D_s^*}$ ratio differ from one. With the exception of scenarios S-VI and S-IX, all the cases show the larger deviations from SM, suggesting the possibility of lepton universality violation in the $b \to s \mu^+ \mu^-$ decay mode. 
    \item With the exception of scenarios S-IV, S-V and S-VI, all other scenarios make significant contributions to the forward-backward asymmetry of the $B_c \to D_s^{*} \mu^+ \mu^-$ decay mode in the low $q^2$ region. 
    \item Except for scenario S-III, the predictions from all other scenarios remain within the SM theoretical uncertainties of the longitudinal polarization asymmetry.
    \item The S-III scenario exhibits a comparatively larger deviation for the $P_1$ observable. 
    \item With the exception of S-IV, S-V, and S-VI, all other scenarios show significant deviations from the Standard Model prediction of $P_2$ observable. 
    \item The S-VIII and S-IX scenarios display maximum sensitivity to the $P_3$ observable.
    \item The predicted $1\sigma$ bands of all scenarios overlap with the SM band for the $P_4^{'}$ observable.
    \item Except for S-IV, S-V, S-VI, S-X and S-XI, all other scenarios show significant deviations from the SM predictions for $P_5^{'}$. 
    \item The S-VIII and S-IX scenarios are particularly sensitive to the $P_8^{'}$ observable. 
\end{itemize}
Before concluding, we briefly comment on the experimental prospects for observing these decay modes. A recent LHCb search using $9\,{\rm fb}^{-1}$ of data established the upper limit $f_c/f_u \times \mathcal{B}(B_c^+\to D_s^+\mu^+\mu^-) < 9.6\times10^{-8}$
at $95\%$ C.L.~\cite{LHCb:2023lyb}. Assuming a statistics-dominated improvement of sensitivity proportional to the inverse square root of the integrated luminosity, the projected reach improves to approximately $4\times10^{-8}$ and $1.7\times10^{-8}$ for integrated luminosities of $50$ and $300\,{\rm fb}^{-1}$, respectively. The branching-ratio predictions obtained in the present work are generally of the order $10^{-8}$--$10^{-7}$ and may therefore become accessible in future LHCb measurements. Improved prospects may also be anticipated for the $B_c\to D_s^{*}\mu^+\mu^-$ mode with the larger datasets expected in future LHCb runs.

\section{Conclusion}
In this work, we investigated the sensitivity of vector and axial-vector coefficients on the $B_c \to D_s^{(*)} \mu^+ \mu^-$ decay mode in the low $q^2\in [1,6]~{\rm GeV}^2$ bin. For this analysis, we updated the allowed ranges for various 1D and 2D coefficients using the latest data on the $b \to s \mu^+ \mu^-$ transitions. Based on the $1\sigma$ range of the 1D and 2D solutions, we estimated the $1\sigma$ predicted values for the branching ratio and lepton non-universality parameter of the $B_c \to D_s \mu^+ \mu^-$ process. Additionally, we analyzed the forward-backward asymmetry, longitudinal polarization asymmetries, and form factor-independent observables for the $B_c \to D_s^{*} \mu^+ \mu^-$ decay in both scenarios. 

Within the framework of 1D scenarios, the coefficients $C_9^{\rm NP} = -C_{10}^{\rm NP}$ and  $C_9^{\rm NP}$ provide a good fit to the $b \to s \mu^+ \mu^-$ and exhibit significant deviations from the SM predictions for all observables in the $B_c \to D_s^{(*)} \mu^+ \mu^-$ decay process. These scenarios also predict $R_{D_s^{(*)}}$ values below one, suggesting potential lepton flavor universality violation in the $b \to s \mu^+ \mu^-$ transition, despite the consistent results for $R_{K^{(*)}}$ reported by the LHCb collaboration. 

From the perspective of 2D scenarios, we observed that the new WC combinations $(C_9^{\rm NP},C_{10}^{\rm NP})$, $(C_9^{\rm NP},C_{10}^{'\rm NP})$,  $(C_{9}^{\rm NP}=C_{9}^{'\rm NP}, C_{10}^{\rm NP}=-C_{10}^{'\rm NP})$  and $(C_{9}^{\rm NP}=-C_{10}^{\rm NP},C_{9}^{'\rm NP}=C_{10}^{'\rm NP})$ provide an excellent fit to all observables in the $b \to s \mu^+ \mu^-$ very well. These 2D scenarios are particularly notable for their sensitivity to the observables of $B_c \to D_s^{(*)} \mu^+ \mu^-$ and exhibit significant deviations from the SM predictions. 

In summary, we have analyzed $B_c \to D_s^{(*)}\mu^+\mu^-$ decays within an effective field theory framework, focusing on the impact of new vector and axial-vector couplings. While the formalism and fitting strategy follow established methods, our study highlights the role of these channels as complementary probes of $b \to s \mu \mu$ transitions. The predictions, though subject to uncertainties from hadronic inputs, provide a baseline for future measurements and offer potential cross-checks of NP scenarios obtained in global fits. A more complete picture will emerge once improved form-factor determinations, LFU-sensitive observables, and experimental prospects are incorporated, which we leave for future work.

\section*{Acknowledgment}
MKM would like to acknowledge the financial support from the IoE PDRF at the University of Hyderabad. AKY expresses gratitude to the DST-Inspire Fellowship division, Government of India, for financial support under ID No. IF210687.   MKM also extends sincere thanks to Dr. Jacky Kumar and Dr. Girish Kumar for their valuable suggestions regarding the flavio package.
\bibliographystyle{apsrev4-1}
\bibliography{referance}
\end{document}